# Monte-Carlo study of the reorientation transition in Heisenberg models with dipole interactions

A. Hucht*, A. Moschel and K. D. Usadel

Theoretische Tieftemperaturphysik

Gerhard-Mercator-Universität –Gesamthochschule– Duisburg

Lotharstraße 1, 47048 Duisburg, Germany

We simulated the classical two-dimensional anisotropic Heisenberg model with full long range dipole interaction with an algorithm especially designed for long range models. The results show strong evidence for a first order reorientation transition at a temperature $T_R < T_C$ for appropriate parameters of the model Hamiltonian.

## Introduction

Recent experiments on thin ferromagnetic films have shown that the magnetic behaviour of these systems is strongly affected by anisotropy fields which act on the spins in the surface [1]. These experiments show that for many systems the direction of the magnetization vector is a very sensitive function of temperature. At low temperatures the anisotropy fields force the magnetization vector to be perpendicular to the surface, but with increasing temperature the magnetization vector switches to an in plane direction. The switching temperature decreases with increasing film thickness. This reorientation transition has been investigated theoretically using classical [2] and quantum mechanical [3] meanfield theory and renormalization group theory [4]. These methods predict a reorientation transition of either first order in classical systems or second order in quantum systems, while a low temperature series expansion shows no evidence for a reorientation transition [5]. These conflicting results call for further investigations.

## The model

Therefore we have performed Monte-Carlo simulations on a classical two-dimensional anisotropic Heisenberg system with long range dipole interactions on a square lattice. The model is described by the Hamiltonian

$$\mathcal{H} = -\frac{J}{2}\sum_{\langle ij \rangle}\mathbf{s}_i\cdot\mathbf{s}_j - K\sum_i (s_i^z)^2 + \frac{D}{2}\sum_{ij} r_{ij}^{-3}\mathbf{s_i}\cdot\mathbf{s_j} - 3r_{ij}^{-5}(\mathbf{s}_i\cdot\mathbf{r}_{ij})(\mathbf{r}_{ij}\cdot\mathbf{s}_j),$$

*email: fred@thp.Uni-Duisburg.DE

where the $\mathbf{s}_i = (s_i^x, s_i^y, s_i^z)$ are classical Heisenberg spins at position $\mathbf{r}_i = (x_i, y_i, 0)$ and $\mathbf{r}_{ij} = \mathbf{r}_i - \mathbf{r}_j$. $J$ is the nearest-neighbour exchange coupling constant, $K$ is the on-site anisotropy, and $D$ is the strength of the long range dipole interaction. We consider ferromagnetic systems with $J = 1$ in this paper.

The ground state energy per spin of a periodic system of fixed size can be calculated exactly as long as $D$ is sufficiently small compared to $J$ to ensure that all spins are aligned parallel. Using the two-dimensional lattice sum [6]

$$\begin{aligned}\Theta(s) &= \sum_{\substack{i,j=-\infty \\ (i,j)\neq(0,0)}}^{\infty} (i^2+j^2)^{-s} \\ &= 4^{1-s}\zeta(s,0)\big(\zeta(s,1/4)-\zeta(s,3/4)\big),\end{aligned}$$

where $\zeta(s,a)$ is the generalized Riemann zeta function and $\mathrm{Re}(s) > 1$, one obtains the ground state energy per spin as a function of the $z$-component of the ground state magnetization $s_{GS}^z$ as

$$e_{GS}(s_{GS}^z) = -2J - K(s_{GS}^z)^2 + \frac{D}{4}(3(s_{GS}^z)^2-1)\Theta\left(\frac{3}{2}\right).$$

Hence all spins are pointing in $z$-direction when $\frac{D}{K} < \frac{4}{3\Theta(3/2)} = 0.1476$, while for $\frac{D}{K} > \frac{4}{3\Theta(3/2)}$ the spins are oriented in an arbitrary direction in the $xy$-plane.

## Method

In the Monte-Carlo simulation we considered quadratic systems with periodic boundary conditions and summed the long range dipole interaction over all periodic images of the system. The simulations were done with an algorithm especially designed for long range systems: The local fields at each site are computed at the beginning of the simulation and are only updated when a spin flip attempt (SFA) is accepted. With this method we obtain a speedup factor of approx. 30 in the relevant temperature regime. To sample the phase space more efficiently, we combine both local and global spin update mechanisms: Single Heisenberg spins are chosen randomly and are reoriented into a random

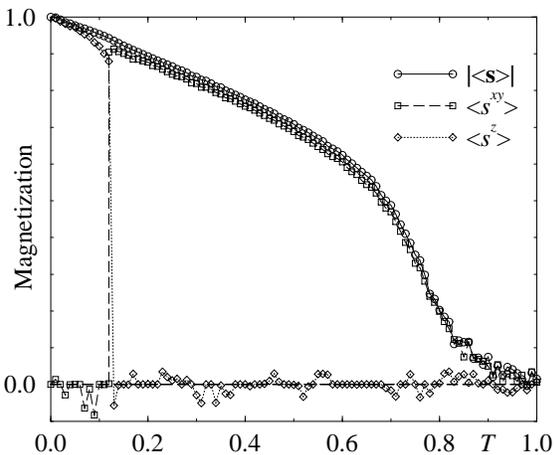

Figure 1: Components of the total magnetization as a function of temperature for a 16 × 16 system with $K = 0.2$, $D = 0.028$

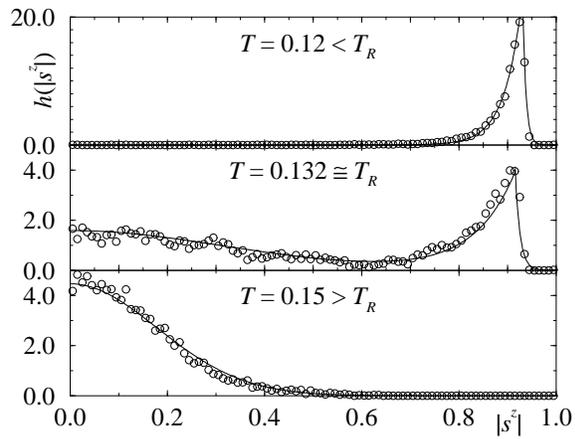

Figure 2: Probability density determined by the simulation $h(|s^z|)$ of the absolute value of the $z$-component of the total magnetization at several temperatures for a 32 × 32 system with $K = 0.2$, $D = 0.028$

direction. Additionally, after a fixed number of local SFAs, a global SFA is made, where all spins are mirrored at a plane with randomly chosen normal vector. Note that this transformation leaves the exchange part of the Hamiltonian invariant. This global spin flip is similar to the cluster spin flip in the Swendsen-Wang Algorithm for Heisenberg models [7], although this algorithm is not applicable in the present model due to the anisotropy of the dipole interaction.

## Results

In figure 1 the perpendicular component $\langle s^z \rangle$, the in-plane components $\langle s^{xy} \rangle$ and the absolute value $|\langle s \rangle|$ of the spontaneous magnetization for a system of 16 × 16 spins are depicted. The values are obtained from the maxima of the distribution function described below. In this run we made 512000/site local and 2000 global SFAs per temperature step using 60 CPU-hours on an IBM RS/6000 590 workstation. As one can see clearly, the magnetization changes from perpendicular to inplane orientation in the temperature range $0.12 < T_R < 0.13$ for this set of parameters.

To investigate the order of the reorientation transition, we considered the probability density

$$h(x) = \lim_{t \to \infty} \frac{1}{t} \sum_{\tau=1}^{t} \delta(x_\tau - x)$$

for $x = |s^z|$, as determined quite accurately from a histogram for a 32 × 32 system at fixed temperature. Here a typical run involved ca. $2 \times 10^6$ local SFAs per site and 7500 global SFAs and took $\approx 10$ CPU-hours on an IBM 590. $h(|s^z|)$ is depicted in figure 2. One can see that the reorientation transition seems to be of first order, as the distribution has two maxima at $T_R$.

## Conclusions

To summarize, we simulated the classical two-dimensional anisotropic Heisenberg model with full long range dipole interaction with an algorithm especially designed for long range models. The first results show strong evidence for a first order reorientation transition at a temperature $T_R < T_C$. This phase transition will be investigated in further detail in the future.

## Acknowledgements

This work was supported by the Deutsche Forschungsgemeinschaft through Sonderforschungsbereich 166.